\documentclass[10 pt, conference]{IEEEtran}
\usepackage{amsmath}
\usepackage{amssymb}
\usepackage{amsthm}
\usepackage{enumerate} 
\usepackage{url}
\usepackage{hyperref}
\usepackage{pgf-umlsd}

\newtheorem{mydef}{Definition}
\newtheorem{myprop}{Property}
\usepackage [lambda,advantage,operators,sets,adversary,landau,probability,notions,logic,ff,mm,primitives,events,complexity,asymptotics,keys]{cryptocode}

\begin{document}
\title{A Practical Scheme for Two-Party Private Linear Least Squares}
\author{\IEEEauthorblockN{Mohamed Nassar}
\IEEEauthorblockA{Department of Computer Science\\
Faculty of Arts and Sciences\\
American University of Beirut\\
Email: mn115@aub.edu.lb}}
\maketitle

\begin{abstract}
Privacy-preserving machine learning is learning from sensitive datasets that are typically distributed across multiple data owners. 
Private machine learning is a remarkable challenge in a large number of realistic scenarios where no trusted third party can play the role of a mediator. The strong decentralization aspect of these scenarios requires tools from cryptography as well as from distributed systems communities.
In this paper, we present a practical scheme that is suitable for a subclass of machine learning algorithms and investigate the possibility of conducting future research.
We present a scheme to learn a linear least squares model across two parties using a gradient descent approach and additive homomorphic encryption.
The protocol requires two rounds of communication per step of gradient descent. We detail our approach including a fixed point encoding scheme, and one time random pads for hiding intermediate results. 
\end{abstract}
\section{Introduction} 

Private machine learning is driven by trust issues in cloud computing. The cloud unlocks unprecedented opportunities for outsourcing of storage and computation. It offers flexibility, scalability  and cost saving, but the risk of being exposed to privacy and security issues retains a lot of customers from risking their sensitive data to the cloud. 
Recent research focuses on leveraging cryptographic techniques to enable 
secure outsourcing of computation to the cloud. 
 Different approaches have been proposed to assimilate this challenge, some of them are based on differential privacy, 
some others are tailored from homomorphic encryption, and some consider the problem as a special case of secure multi-party computation and secure function evaluation.

A cryptosystem which supports both addition and multiplication (thereby preserving the ring structure of the plaintexts) is known as Fully Homomorphic Encryption (FHE). FHE effectively allows the construction of programs which may be run on encryptions of their inputs to produce an encryption of their output. Since such a program never decrypts its input, it can be run by an untrusted party without revealing its inputs and internal state. This would have great practical implications in the outsourcing of or multi-party secure computation. 

Gentry describes a solution to fully homomorphic encryption based on Lattice cryptography \cite{gentry2009fully}. Unfortunately it is estimated that performing a Google search with encrypted keywords would increase the amount of computing time by about a factor of trillion. This has been later improved through several optimizations. Currently, HELib (\url{https://github.com/shaih/HElib}) evaluates the arithmetic circuit of an AES-128 block in around 2 seconds, amortized time. 

Somewhat homomorphic encryption (SHE) schemes, which support a limited number of homomorphic operations, can be much faster, and more compact than fully homomorphic encryption ones. Nevertheless, noise grows exponentially with respect to the number of levels of multiplications performed. Bootstrapping, a very costly procedure, is then necessary to reduce the noise to its initial level. Leveled homomorphic encryption (LHE) allows evaluation of polynomial functions of a bounded degree without resorting to bootstrapping. 
  
\begin{figure}
    \centering
    \includegraphics[width=\linewidth]{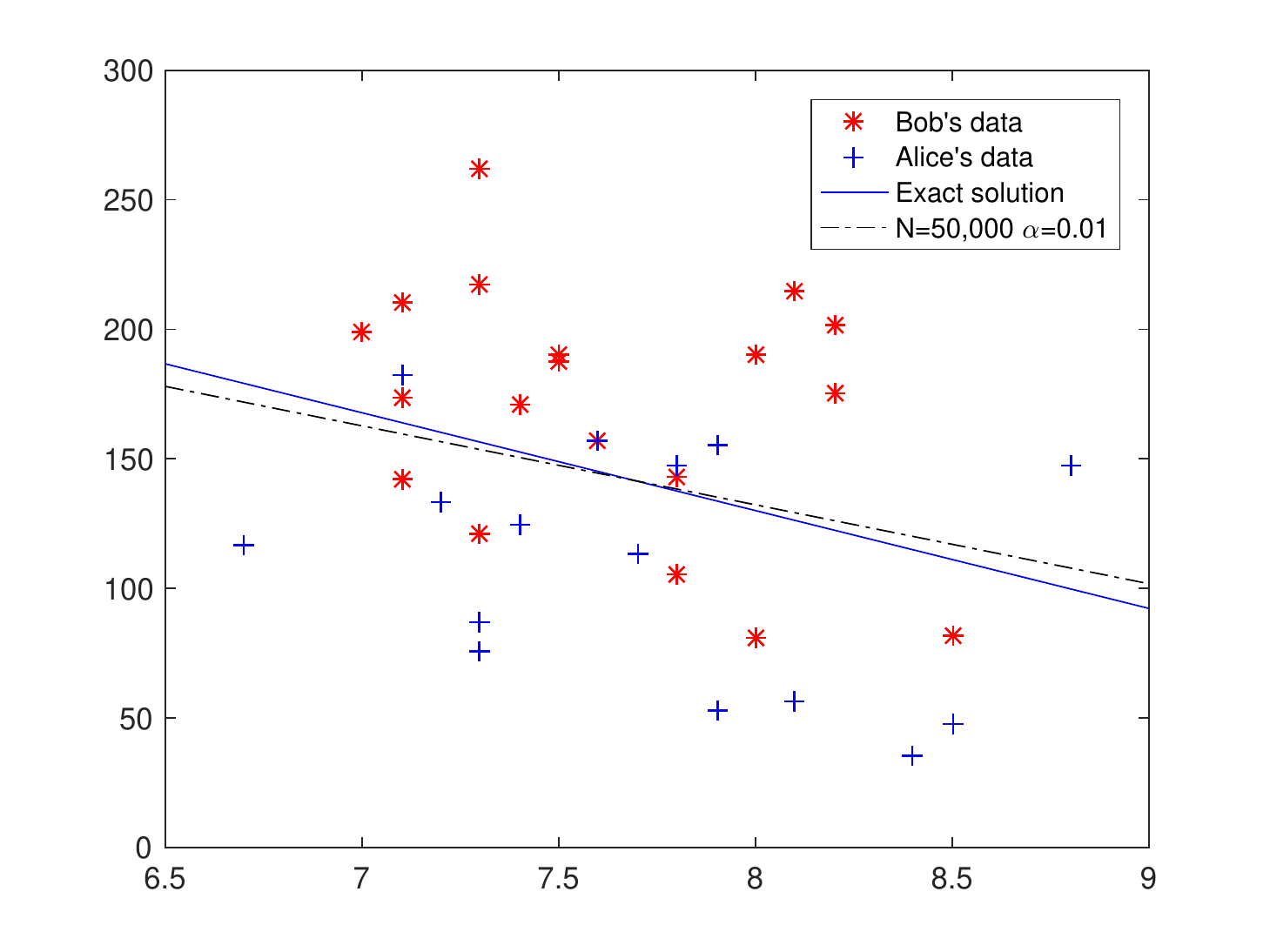}
    \caption{Two-Party Linear Regression: Exact vs. Gradient Descent Solutions}
    \label{gradient_descent}
\end{figure}

Partially homomorphic cryptosystems are simpler and support only one kind of computation (e.g. addition, multiplication, XOR). However they are more practical in terms of performance and have a wide range of applications ranging from secure voting and collision resistant hash functions to private information retrieval and secure computation on the cloud. Following the seminal work of Atallah et. al. \cite{atallah2002secure}, we use Paillier's additive homomorphic encryption \cite{Paillier1999} to provide a practical solution for several secure distributed learning problems. We focus on a subclass of machine learning algorithms called D-polynomial as per the definition in \cite{graepel2012ml}. We consider in particular two-party private linear least squares and linear regression. The proposed schemes can be extended to the multi-party scenario. Figure \ref{gradient_descent} shows two solutions for linear regression  based on data incoming from two parties: Bob and Alice. The first solution is the exact solution, while the second is based on the method of gradient descent. However, the gradient descent converges to the exact solution if we iterate long enough and choose a small enough learning rate. The goal of this paper is to let the two parties compute the solution to any linear least squares under privacy constraints by the means of a distributed gradient descent protocol. 

The rest of the paper is organized as follows: Section II summarizes related word. We present background information about linear least squares in Section III. Section IV presents a summary of the homomorphic cryptosystem that we are using, problem formulation and threat model. We propose the private machine learning protocol in Section V. We discuss simulation results and implementation choices in Section VI and conclude the paper in Section VII. 

\section{ Related Work } 
We divide related work into two categories: 
\begin{itemize} 
 \item Work that addresses secure outsourcing of computation. This work stresses on the asymmetry of computation workload. In other words,  
  computationally weak clients are the owners of the data and want to outsource their data processing tasks to a computationally strong server.  
 This work is more suitable for the context of cloud computing. The survey in \cite{shan2018practical} distinguishes between two different applications: 
 \begin{itemize} 
 \item Fundamental functions such as scalar operations, e.g. Exponentiation modulo a large integer for RFID tags \cite{chen2014new}, matrix operations \cite{benjamin2008private} and mathematical optimization \cite{wang2011secure}. 
 \item Application-oriented procedures such as machine learning \cite{graepel2012ml}, image processing \cite{wang2013secure}, and biometric authentication \cite{blanton2010secure}
 \end{itemize}
  \item The second category, like in this paper, addresses multi-party secure function evaluation. We are especially interested in machine learning functions. Given $m$ instances in $\mathbb{R}^n$, their categorical or numerical labels in 
$\mathcal{Y}$ and an unlabeled instance in $\mathbb{R}^n$, the function outputs a label in $\mathcal{Y}$:  $$A: (\mathbb{R}^n \times \mathcal{Y})^m \times \mathbb{R}^n \rightarrow \mathcal{Y}$$
In this context, the data is split horizontally or vertically in between different parties.  The different parties are willing to learn a machine learning model over the overall data set without unveiling their owned data parts. 
 We put in this category the work in \cite{du2004privacy}, \cite{du2001privacy}, \cite{karr2005secure} and \cite{hall2011secure}
\end{itemize}

We also can categorize related work based on the privacy preserving strategy in question: 
\begin{itemize} 
 \item work that are based on cryptography, data encryption using homomorphic encryption or Yao's garbled circuits \cite{yao1982protocols}. Our paper falls in this category; 
 \item and work based on data anonymization by adding special types of statistical noise to the data, such as in differential privacy. \cite{zhang2012functional}. 
\end{itemize}

In \cite{graepel2012ml}, leveled homomorphic encryption scheme is proposed to delegate the execution of a machine learning algorithm to a computing service while retaining confidentiality of the training and test data. Since
the computational complexity of the homomorphic encryption scheme depends primarily on the number of levels of multiplications to be carried out on the encrypted data, a new class of machine learning algorithms is defined in which the algorithm’s predictions, viewed as functions of the input data, can be expressed as polynomials of bounded degree. For instance, comparison and division are not polynomial. 
We take linear means classifier as an example to depict their methodolgy. 
The linear means classifier defines a hyperplane midway on and orthogonal to the line through the two class-conditional means. This can be derived as the Bayes optimal decision boundary in the case that the two class-conditional distributions have identical isotropic distributions. 
Formally, the classification model is defined by a vector $\theta$ and an offset $c$. 
$$ \theta = M_1 - M_2 $$ 
$$ c = \theta^T\frac{(M_1 + M_2)}{2}$$
where $ M_1 = \frac {S_1}{m_1} $ is the mean of all vectors belonging to $C_1$ and $M_2 = \frac {S_2}{m_2} $ is the mean of all vectors belonging to $C_2$. $m_1$ and $m_2$ are the respective count of sets $C_1$ and $C_2$.  $S_1$ and $S_2$ are their respective vectors sum.
 
The prediction of a new vector $x$ depends on the sign of the score function $ A(\theta, c, x) = \theta^T x - c$, if positive the class is $class1$ and if negative the class is $class2$.

A division-free form of the same prediction can be obtained through multiplication by $2m_1^2m_2^2$: 
\begin{align*}
A_{\text{divsion-free}}(\theta, c, x) = 2m_1^2m_2^2  ( \frac{S_1^T}{m_1} - \frac{S_2^T}{m_2} )   ( x -  \frac{S_1}{2m_1} - \frac{S_2}{2m_2} ) \\
= (m_2 S_1^T - m_1 S_2^T ) ( m_1m_2x - m_2 S_1 - m_1 S_2  ) 
\end{align*} 

Linear means classifiers ($\mathbf{\theta}^T\mathbf{x} - c=0$) is 2-Polynomial because $c$ is a quadratic function in terms of the training data vector. Division-free integer algorithms are proposed as approximations to solve binary classification and  least-squares using a small number of gradient descent steps. We follow a very similar approach in this paper, although our schemes only use PHE without recurring to LHE. 

An interesting approach for practical secure aggregation
with the goal of private machine learning is proposed in \cite{bonawitz2017practical}. A distributed system with crash node failures and malicious node failures is presented. 
Distributed systems protocols are designed to allow collecting and computing the sum of large, user-held data vectors incoming from mobile devices at a central server without the server learning individual users contribution.
The individual users contributions are protected using one-time masking pads.  The collected data can be used in a federated learning setting, to train a deep neural network. We also use one time masking pads in this paper.

\subsection*{SMC Implementations}
Several frameworks are proposed in the literature for implementing secure multi-party computation. The most important ones are Obliv-C, ObliVM, SPDZ and Sharemind \cite{doerner}. Obliv-C (\url{https://oblivc.org}) implements Yao's Garbled Circuits and other techniques in a C-compatible domain specific language. Obliv-C achieves minimalism and expressiveness through a few additional keywords to the C language such as \texttt{obliv} and oblivious functions (e.g. \texttt{feedOblivInt()}, \texttt{revealOblivInt()}). The authors report a raw speed of 3M+ AND gates per second. ObliVM (\url{http://oblivm.com}) follows the same philosophy and its language has a Java/C++ style. It reports 700K AND gates per second
or 1.8M with preprocessing. 
More interesting to our approach is SPDZ (\url{http://www.bris.ac.uk/engineering/research/cryptography/resources/spdz-software/}) which support SHE and linear secret sharing. It does not have a domain specific language but can be programmed via Python library calls. It reports speeds in the range from 4800 multiplications/second for two parties in the offline phase to 358K multiplications/second for two parties in the online phase. 
Finally Sharemind (\url{https://sharemind.cyber.ee}) is a commercial application Platform but available for researchers. Its style is similar to Java or .NET and has unique features such as vector optimization.  We are to evaluate these tools for experimenting with private machine learning in general and our proposed protocols in particular.

\section{Background on Linear Least Squares} 
We start by recalling essential background information on linear least squares.

In least squares (and linear regression in particular), we have a number $m$ of data items represented in a $n$ dimensional space. The data items form a long and skinny $m\times n$ matrix $X$ and their corresponding values form a vector $Y\in \mathbb{R}^m$. We assume that $X$ has a full column rank. In the over-determined case  $m > n$, there is typically no vector $\theta \in \mathbb{R}^n $ satisfying $X\theta=Y$. 

\subsection{Normal equations} 
We aim to find a vector $\theta$ that minimizes the loss function  $J= \frac{1}{2}||X\theta - y||^2$. 
Let's first compute the gradient of $J$ with respect to $\theta$:
\begin{align}
\notag J(\theta)  & = \frac{1}{2} \Vert X\theta - y \Vert ^2 \\
\notag & =  \frac{1}{2}(X\theta - y)^T (X\theta - y) \\
\notag & = \frac{1}{2}( \theta^T X^T - y^T ) ( X\theta - y )  \\
\notag & = \frac{1}{2} (\theta^T X^TX \theta - 2 y^T X \theta + y^T y ) 
\end{align} 
Taking the gradient with respect to $\theta$ gives: 
$$ \frac {\partial J(\theta) } {\partial \theta} = X^T X \theta - X^T y $$
The minimum can be found directly by solving the system of so-called normal equations $ X^T X \theta - X^T y = 0 $: 
$$ \theta = (X^TX)^{-1} X^Ty $$ 
However, we cannot afford the matrix inversion and the matrix multiplications under encryption. We solve for $\theta$ using a gradient descent method as described next.  

\subsection{Gradient descent}
The gradient descent method consists on starting from a random guess of $\theta$, say $\theta_0$ (where values are usually chosen small), 
and consecutively update $\theta$ through many iterations. Each update walks a small step as controlled by a learning rate parameter $\alpha$ where 
$0<\alpha<1$,  
in the opposite direction of the gradient of the loss function.

The formula to update the vector $\theta$ is as follows: 
$$ \theta_{i+1} = \theta_{i} - (\alpha/m) (X^T (X\theta - y ) ) $$

\section{Partially Homomorphic encryption}
There is no universal method to create a protocol for secure multi-party computation.
 Several homomorphic systems only support a subset of mathematical operations, like addition (Paillier, Benaloh), multiplication (ElGamal, RSA), or exclusive-or (Goldwasser and Micali). From a security perspective, only the additive Paillier and the multiplicative ElGamal are classified to be IND-CPA (stands for indistinguishability under chosen plaintext attack) \cite{fontaine2007survey}. Partially homomorphic cryptosystems are more desirable from a performance point of view than somewhat homomorphic cryptosystems, which support a limited operation depth. 
 \subsection{Paillier's cryptosystem}
We use Paillier's homomorphic cryptosystem \cite{Paillier1999} that possesses the following properties:
\begin{enumerate}[(i)]
\item It's a public key scheme, which means encryption can be performed by anyone who knows the public key, whereas decryption can only be done by the matching private key, known only to a trusted party. 
\item It is probabilistic. In other words, it is impossible for an adversary to tell whether two ciphertexts are encryptions of the same plaintext or not. 
\item  It possesses the homomorphic properties for addition, in particular:
\end{enumerate}
\begin{align}
\mathcal{E}_{\pk}[(m_1+m_2) \mod N_{\pk}] &\equiv \mathcal{E}_{\pk}[m_1]\mathcal{E}_{\pk}[m_2] \mod N_{\pk}^2 \\ 
\mathcal{E}_{\pk}[(a.m_1)\mod N_{\pk}] &\equiv \mathcal{E}_{\pk}[m_1]^a  \mod N_{\pk}^2 
\end{align}
Where $N_{\pk}=pq$ is part of the public key.
We can build operations over matrices and vectors on top of these properties:
\begin{itemize}
\item addition of two matrices (or vectors) under encryption, 
\item matrix multiplication of an encrypted matrix or vector by a plaintext matrix or vector.
\end{itemize}

\subsection{Problem Formulation and Threat Model}
Our threat model is "honest but curious", or as often called "semi-honest". We assume that the parties run the protocol exactly as specified, therefore we do not assume any deviation from the protocol, malicious or other. However, the parties may try to learn as much as possible about the input of the other party from their views of the protocol exchanged data. Hence, we want the view of each party not to leak more knowledge than the prior knowledge, as formally expressed in the following definition:

\begin{mydef} 
\label{defsec}
A pair of probabilistic polynomial-time Turing machines $(P_1, P_2)$ is a secure 2-party protocol (for static, semi-honest adversaries) for a deterministic polynomial time-computable function $f$ if the following properties hold:
\begin {itemize} 
\item Completeness: for all $i \in \{1, 2\}$ and inputs $x_1, x_2 \in \{0,1\}^*$, we have (with probability 1): $$ out_{P_i}[P_1(x_1) \leftrightarrow P_2(x_2) ] = f(x_1, x_2) $$ 
\item Privacy: there exist non-uniform probabilistic polynomial-time simulators $\mathcal{S}_1, \mathcal{S}_2$ such that for all $x_1, x_2 \in \{0, 1\}^*$ and all $i \in \{0,1\}$: 
$$ view_{P_i}[P_1(x_1) \leftrightarrow P_2(x_2) ] \overset{c}{\approx} S_i(x_i, f(x_1, x_2) ) $$ 
\end {itemize} 
\end{mydef}

To cope with the practical aspects of Definition \ref{defsec} we derive the following property that we call ``safe exposure'':  
\begin{myprop}
A data item $x_i$ that belongs to a party $P_i$ has a ``safe exposure''  during and following a protocol run 
if it is not exposed to any other party, except through one of these three  means: 
\begin{itemize} 
 \item the item is encrypted using partially homomorphic encryption, the private key of which belongs to the owner party $P_i$. 
 \item the item is added to a securely generated random number $R$ and exposed as $X+R$ where $R$ is only known to the owner party $P_i$.
 \item the item is used in a proved secure multi-party computation, particularly a garbled-circuits based protocol.  
\end{itemize}
\end{myprop}

An alternative and equivalent definition can be expressed as follows: 
\begin{mydef}
A two-party protocol is secure with respect to a problem definition, if the safe exposure property if enforced for all tuples $(x_i, P_i)$ 
where data item $x_i$ is required to be private to party $P_i$ in the problem definition. 
\end{mydef}
\begin{figure*}[htbp]
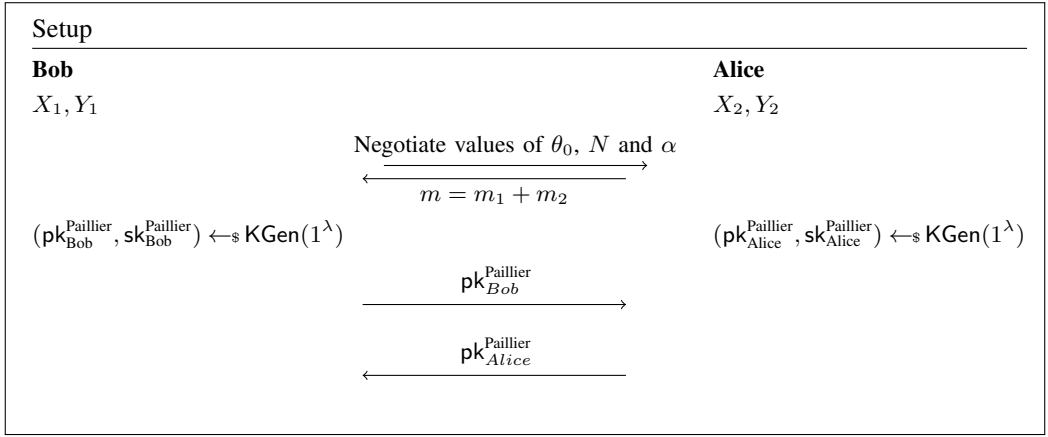

\centering
\fbox{
\procedure{Setup}{
\textbf{Bob} \< \<  \textbf{Alice} \\ 
X_1, Y_1 \< \<  X_2, Y_2 \\
\< \sendmessageright{length=3.5cm, top={Negotiate values of $\theta_0$, $N$ and $\alpha$}} \<\\[-0.8\baselineskip]
\< \sendmessageleft{length=3.5cm, bottom={$m = m_1 + m_2$}} \\
(\pk_{\text{Bob}}^{\text{Paillier}}, \sk_{\text{Bob}}^{\text{Paillier}}) \sample \kgen (\secparam) \< \< (\pk_{\text{Alice}}^{\text{Paillier}}, \sk_{\text{Alice}}^{\text{Paillier}}) \sample \kgen (\secparam)\\ 
\< \sendmessageright*[3.5cm]{\pk_{Bob}^{\text{Paillier}}} \<\\
\< \sendmessageleft*[3.5cm]{\pk_{Alice}^{\text{Paillier}}} \<\\
}
}
\caption{Setup phase of the protocol}
\label{setup1}
\end{figure*}

\section {Privacy Preserving Protocols}
Bob and Alice want to learn a model of their respective data sets using gradient descent while preserving the privacy of their data.  
Bob has one part of the data $X_1$ of cardinal $m_1$ and labels $Y_1$ and Alice has another part $X_2$ of cardinal $m_2$ and labels $Y_2$. 
The concatenation of their data is $X$. Each row of $X$ represents the feature vector of one data item belonging to Alice or Bob. In the setup phase, Bob and Alice exchange their public keys for Paillier's homomorphic encryption given a security parameter $\lambda$.  Bob and Alice negotiate values for the learning rate $\alpha$, number of iterations to run $N$ and the starting value $\theta_0$. They also compute the total number of instances $m$ as shown in Figure \ref{setup1}.  

To solve the problem of linear least squares we use 
the gradient descent method. Basically it consists on  initializing a random vector $\theta_0$ with random values (usually $\ll 1$), 
a learning rate $\alpha$, then run a sufficient number of iterations $N$ to successively move $\theta$ a small step in the opposite direction of the gradient. 
The update formula is as follows: 
$$ \theta_1  = \theta_0 - (\alpha/m) ( X^TX\theta_0 - X^TY ) $$ 
where $X$ is the feature matrix of dimension $m \times n$, $m$ is the number of data items and $n$ is the number of features. In the case of linear regression $n-1$ is the degree of the polynomial representing the solution of the least squares optimization. 

In the two-party model, the matrix $X$ is composed of two parts: $X_1$ is Bob's data and $X_2$ is Alice's data.
$$ X = \left ( \begin{array}{c}  X_1 \\ X_2 \end{array}  \right) $$ 
$$ Y = \left ( \begin{array}{c}  Y_1 \\ Y_2 \end{array}  \right) $$ 
$$ X^TX =
\left ( \begin{array}{cc}  X_1^T \ X_2^T  \end{array}  \right)
\left ( \begin{array}{c}  X_1 \\ X_2 \end{array}  \right)
= 
X_1^TX_1 + X_2^TX_2 
$$
$$
X^TY = \left ( \begin{array}{cc}  X_1^T \ X_2^T  \end{array}  \right)
\left ( \begin{array}{c}  Y_1 \\ Y_2 \end{array}  \right)
= 
X_1^TY_1 + X_2^TY_2 
$$
Note that Bob can compute $X_1^TX_1$ and $X_1^TY_1$, and Alice can compute $X_2^TX_2$ and $X_2^TY_2$ independently and without any data transfer. 

Bob and Alice agree on a fixed point representation to encode real numbers into integers (since our encryption scheme only works with integers). 
We choose a scaling factor $s>0$ in function of the fractional digits needed. A real number $r$ is mapped into an integer $t$ as follows: 
$$ t = \text{round}( r * 10^s )  $$ 
The decoding of $t$ is the inverse operation yielding $r'$: 
$$ r' = t * 10^{-s}  $$ Therefore the absolute encoding error is $$ |r'-r| < \frac{10^{-s}}{2} $$  
To encode $X^TX\theta_0 - X^TY$, we choose different encoding for elements of $X$, $Y$ and $\theta$. 
If we encode elements of $X$ with a scaling factor $s_1$, the elements of $\theta$ with a scaling factor $s_2$, then the elements of the matrix multiplication 
$X^TX\theta_0$ is encoded with a scaling factor $2s_1 + s_2$. 
To have the same encoding for $X^TY$, elements of $Y$ must be encoded with a scaling factor $s_1+s_2$. The result $X^TX\theta_0 - X^TY$ has a scaling factor of $2s_1 + s_2$.  
Based on this scheme, we use the terms 'encrypt' and 'decrypt' as shorts for 'encode and encrypt', and 'decrypt and decode'. Our protocol for one step of the gradient descent is depicted in Figure \ref{onestep}.  

\begin{figure*}[htbp]
\centering
\fbox{
\procedure{Gradient Descent}{
\textbf{Bob} \< \<  \textbf{Alice} \\ 
X_1^T X_1, X_1^TY_1 \< \<  X_2^T X_2, X_2^T Y_2 \\
\< \sendmessageright*[5cm]{\mathcal{E}_B(X_1^TX_1), \mathcal{E}_B(X_1^TY_1)} \< \\
\< \< \text{Compute:} \\
\< \< \mathcal{E}_B(X_1^T X_1 + X_2^T X_2) = \mathcal{E}_B(X^T X)\\
\< \< \mathcal{E}_B(X_1^T Y_1 + X_2^T Y_2) = \mathcal{E}_B(X^T Y)\\
\< \< \text{Pick: } R_0^A \\ 
\< \< \text{Compute: }\mathcal{E}_B(X^T X \theta_0 - X^T Y + R_0^A) \\
\< \sendmessageleft*[5cm]{\mathcal{E}_B(X^T X \theta_0 - X^T Y + R_0^A)} \<\\
\text{Decrypt: } X^T X \theta_0 - X^T Y + R_0^A \< \< \\
\text{Compute: } \< \< \\ 
\theta_0 - \frac{\alpha}{m}(X^T X \theta_0 - X^T Y + R_0^A) = \< \< \\
\theta_1 - \frac{\alpha}{m}  R_0^A \< \< \\
\text{Pick: } R_0^B \< \< \\
\< \sendmessageright*[5cm]{\theta_1 - \frac{\alpha}{m}  R_0^A +R_0^B} \<\\
\< \< \text{Remove: } -\frac{\alpha}{m}  R_0^A \\
\< \< \text{Obtain: } \theta_1 + R_0^B \quad \textbf{(1)}\\
\< \< \text{Compute: } \\ 
\< \< \mathcal{E}_B(X^TX \theta_1 + X^TX R_0^B -X^TY)\\
\< \< \text{Pick: } R_1^A\\
\< \sendmessageleft*[5cm]{\mathcal{E}_B(X^TX \theta_1 + X^TX R_0^B -X^TY + R_1^A)} \<\\
\text{Decrypt: } \< \< \\
X^TX \theta_1 + X^TX R_0^B -X^TY + R_1^A = \< \< \\
X^TX \theta_1 + X_1^TX_1 R_0^B + X_2^TX_2 R_0^B -X^TY + R_1^A \< \< \\
\text{Remove: } X_1^T X_1 R_0^B \< \< \\
\text{Obtain: } \< \< \\
X^TX \theta_1 + X_2^TX_2 R_0^B -X^TY + R_1^A \quad \textbf{(2)} \< \< \pclb
\pcintertext[dotted]{Sub Protocol for removing $X_2^TX_2 R_0^B$ (Figure \ref{remove}) }\pclb
[-0.8\baselineskip]
\pcintertext[dotted]{End Sub Protocol}\\
\text{Obtain: } X^TX \theta_1 -X^TY + R_1^A -R_2^A \< \< \\
\text{Compute: } \<\< \\ 
\theta_1 - \frac{\alpha}{m}  R_0^A - \frac{\alpha}{m} (X^TX \theta_1 -X^TY + R_1^A -R_2^A) \< \< \\
= \theta_2 - \frac{\alpha}{m} (R_1^A -R_2^A+R_0^A) \< \<  \\
\< \sendmessageright*[5cm]{\theta_2 - \frac{\alpha}{m} (R_1^A -R_2^A+R_0^A) + R_2^B} \<\\
\< \< \text{Obtain: } \theta_2 + R_2^B \\ 
\< \<\text{ Go To } \textbf{(1)} \\
}
}
\caption{Protocol for one-step gradient descent}
\label{onestep}
\end{figure*}

\begin{figure*}[htbp]
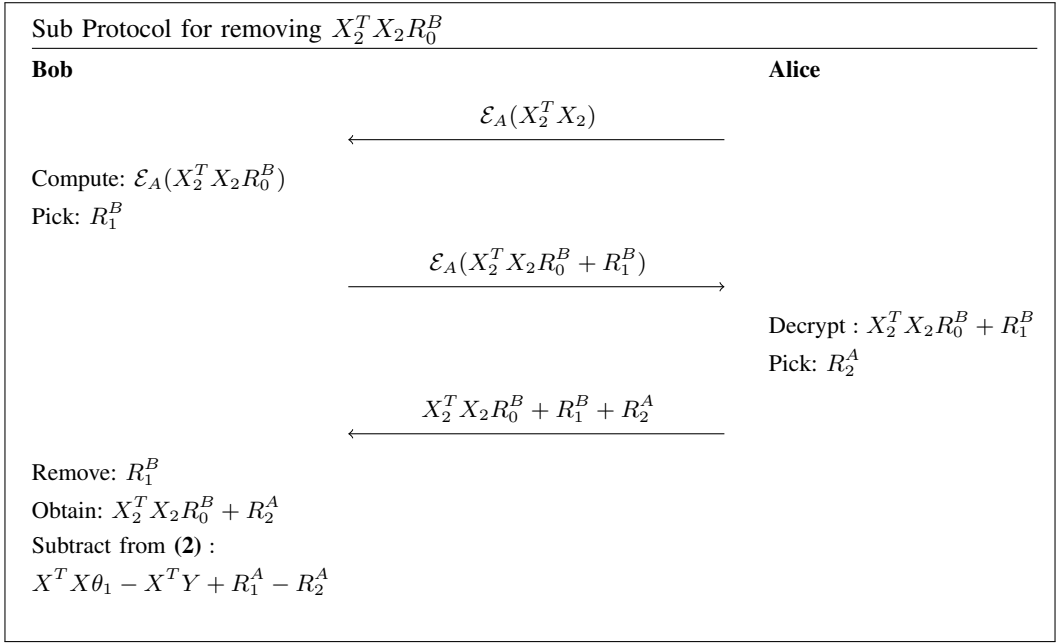

\centering
\fbox{
\procedure{Sub Protocol for removing $X_2^TX_2 R_0^B$}{
\textbf{Bob} \< \<  \textbf{Alice} \\
\< \sendmessageleft*[5cm]{\mathcal{E}_A(X_2^TX_2)} \< \\
\text{Compute: } \mathcal{E}_A(X_2^T X_2 R_0^B) \< \< \\
\text{Pick: } R_1^B \< \< \\
\< \sendmessageright*[5cm]{\mathcal{E}_A(X_2^T X_2 R_0^B+R_1^B)} \< \\
\< \< \text{Decrypt}: X_2^T X_2 R_0^B+R_1^B\\
\< \< \text{Pick: } R_2^A\\
\< \sendmessageleft*[5cm]{X_2^T X_2 R_0^B+R_1^B+R_2^A} \< \\
\text{Remove: } R_1^B \< \< \\
\text{Obtain: } X_2^T X_2 R_0^B+R_2^A \< \< \\
\text{Subtract from \textbf{(2)}}:  \< \< \\ 
X^TX \theta_1 -X^TY + R_1^A -R_2^A \< \< \\
}
}
\caption{Sub-Protocol for one-step gradient descent}
\label{remove}
\end{figure*}

Bob encrypts $X_1^TX_1$ using the additive homomorphic cryptosystem with Bob's public key and sends the encrypted result to Alice. Using the same key, Alice computes $\mathcal{E}_B(X_2^TX_2)$. Based on additive homomorphic encryption properties, Alice computes 
$$ \mathcal{E}_B(X^TX) =  \mathcal{E}_B(X_1^TX_1 + X_2^TX_2) $$
Similarly Alice encrypts $X_2^TY_2$ and computes: 
 $$  \mathcal{E}_B(X^TY) =  \mathcal{E}_B(X_1^TY_1 + X_2^TY_2) $$
Next Alice multiplies by plain text $\theta_0$ to get the result: 
 $$  \mathcal{E}_{B}(X^TX\theta_0 - X^TY) $$
Alice adds a vector of random numbers to the encrypted result: 
$$  \mathcal{E}_B(X^TX\theta_0 - X^TY +R^A_0) $$ 
Alice sends the result back to Bob. Bob decrypts the result. Bob knows nothing about Alice's data since it is masked by $R_0^A$. 
Bob computes $ \theta_0- (\alpha/m) (  X^TX\theta_0 - X^TY +R_0^A ) = \theta_1 - (\alpha/m) R_0^A $. The result is masked by a random vector $R^B_0$ and sent back to Alice. 
Alice removes $- (\alpha/m) R_0^A $ and gets $\theta_1 +R_0^B$. In this way Alice doesn't discover $\theta_1$ and can't use it to recover anything about Bob's data. 

Alice repeats the same procedure as with $\theta_0$ resulting in $ \mathcal{E}(X^TX\theta_1 - X^TY + X^T X R_0^B)$. The difference is that we get an additional term $X^T X R_0^B$ that needs to be removed. Alice picks a random vector $R_1^A$ and add it to the encrypted term sent to Bob.

Since $X^T X R_0^B = X_1^T X_1 R_0^B + X_2^T X_2 R_0^B$, Bob decrypts and removes $X_1^T X_1 R_0^B$. We still need to remove $X_2^T X_2 R_0^B$. To this end, Bob and Alice run the sub-protocol in Fig. \ref{remove} using Alice's generated key this time and only three network messages. The outcome of this sub-protocol is Bob getting $X^TX\theta_1-X^TY$ masked by random vectors belonging to Alice. Bob and Alice proceed to compute $\theta_2$ at Alice's side but masked with another Bob's random vector. 
This ends the first two iterations of our protocol. More iterations can be performed in the same manner leading to $\theta_N + R_N^B$ after $N$ iterations. Each iteration costs five network messages and simple local computations at both sides. The two parties agree on revealing $\theta_N$ simply by having the first party publish $R_N^B$.

The unknowns of the system for Alice are $\theta_{N-1}$, $X_1^TX_1$ and $X_1^TY_1$ 
totaling $n^2 + 2n$ unknowns for only $n$ equations as determined by the following system:  
$$ \theta_N  = \theta_{N-1} - \frac{\alpha}{m} ( (X_1^TX_1 + X_2^TX_2)\theta_{N-1} - (X_1^TY_1 + X_2^TY_2) ) $$
The same applies for Bob. Therefore, the leakage is limited to what can be inferred from a party's data and the result of the linear regression solution.

\section{Simulation Results}
We evaluate our protocol for numerical stability since we are transforming floating point numbers into integers or big integers and back. We also study the time and communication cost. The code for a toy example is available at \url{https://github.com/mnassar/private-two-party-leastsquares}.

\subsection{Numerical Stability} 
\begin{figure}
\centering
\includegraphics[width=0.51\textwidth]{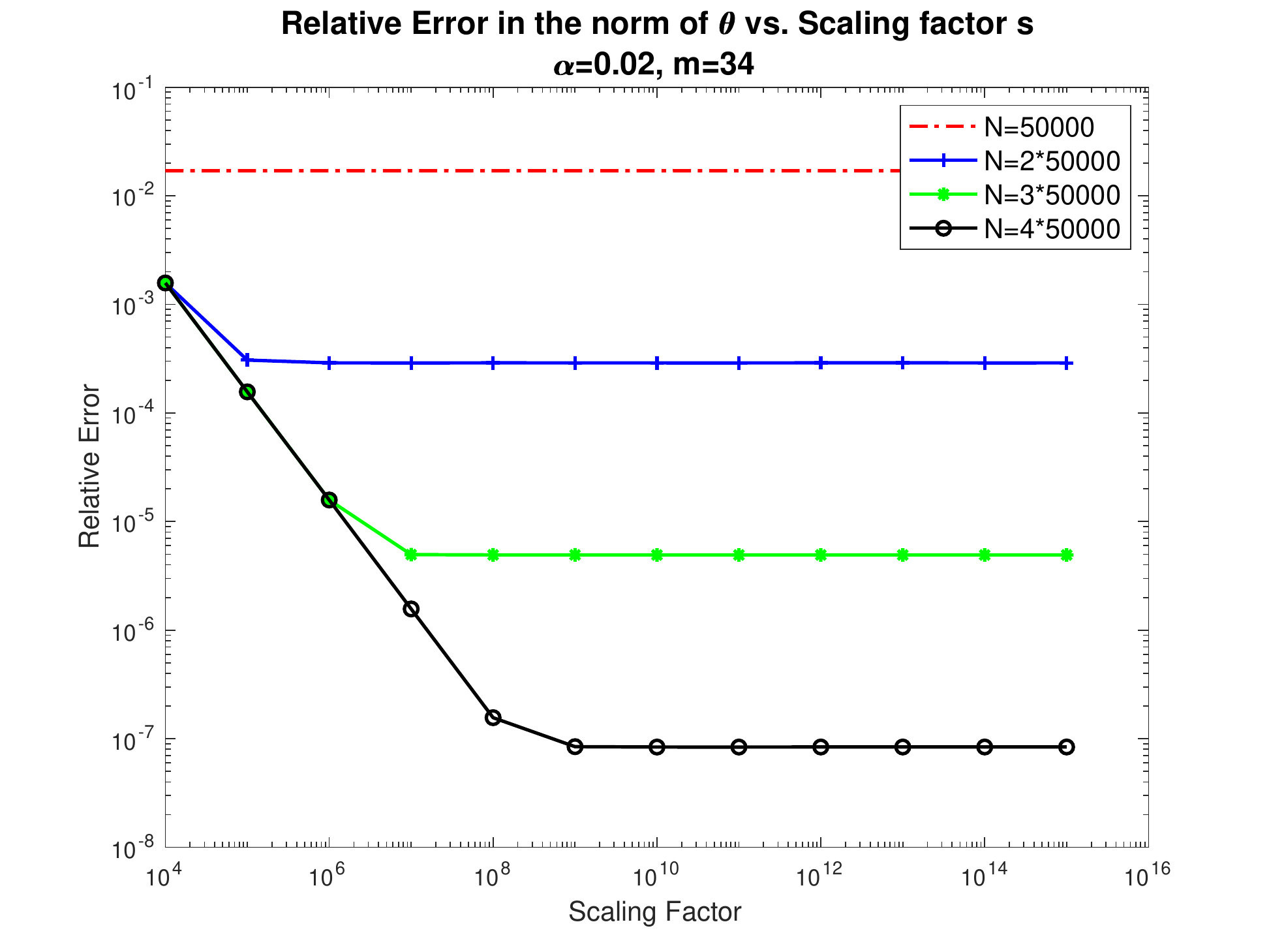}
\caption{Relative Error in $\theta$ vs. scaling factor}
\label{fig:my_label}
\end{figure}

In this experiment we study the effect of the scaling factors $s_1$ and $s_2$ on the relative error in the norm of the solution $\theta$ with respect to the exact solution. Since encryption and decryption are not lossy, the error is due to the normalization and denormalization process.
For simplicity we take the case where $s_1=s_2=s$. We actually have two sources of errors: the convergence error and the rounding error. Figure \ref{fig:my_label} shows that the relative total error is tolerable if we choose good parameters (learning rate, number of required iterations) coupled with a large enough scaling factor. For $N=50,000$ the relative error is independent of the scaling factor since the protocol did not have enough time to converge. For all other experiments, we get a smaller error by increasing the number of iterations and the scaling factor simultaneously. A relative error of $10^{-7}$ is quite tolerable after 200,000 iterations. This is comparable to rounding errors that usually accompany any numerical computations in double precision. These results are consistent among many real world datasets that we have tried. The reason is that most of the time the elements of the matrix $X$ does not have a big difference in scale. We do not encounter numerical instabilities as long as $X$ is not especially crafted to be ill conditioned.

\subsection{Time Cost} 
We have implemented a high performance Paillier's cryptosystem in Python with a GMP C backend (\url{https://github.com/mnassar/paillier-gmpy2}) \cite{nassar2015paillier}. 
Based on our benchmark of the tool (run on one processor Intel Xeon CPU 2.90GHz on a Linux server machine), the average time required for each cryptographic operation are as follows for a key size of 1024 bits: 

\begin{itemize}
 \item Encryption (E): 1000 operations take $\sim 2.1$ seconds.
 \item Decryption (D): 1000 operations take $\sim 1.9$ seconds.
 \item Add plaintext constant to cipher (i.e. an encryption followed by a modular  multiplication (AC)): 1000 operations take $\sim 1.9$ seconds.
 \item Multiply number by cipher (i.e. a modular exponentiation operation (ME)): 1000 operations take $\sim 1.9$ seconds.
 \item Flipping the sign of cipher (a modular multiplicative inversion (MI)): 10,000 operations take $\sim 0.2$ seconds.
  \item Adding two ciphers (i.e. modular multiplication (MM)): 10,000 operations take $\sim 0.05$ seconds.
\end{itemize} 

\begin{figure}[t]
\centering
\includegraphics[width=0.51\textwidth]{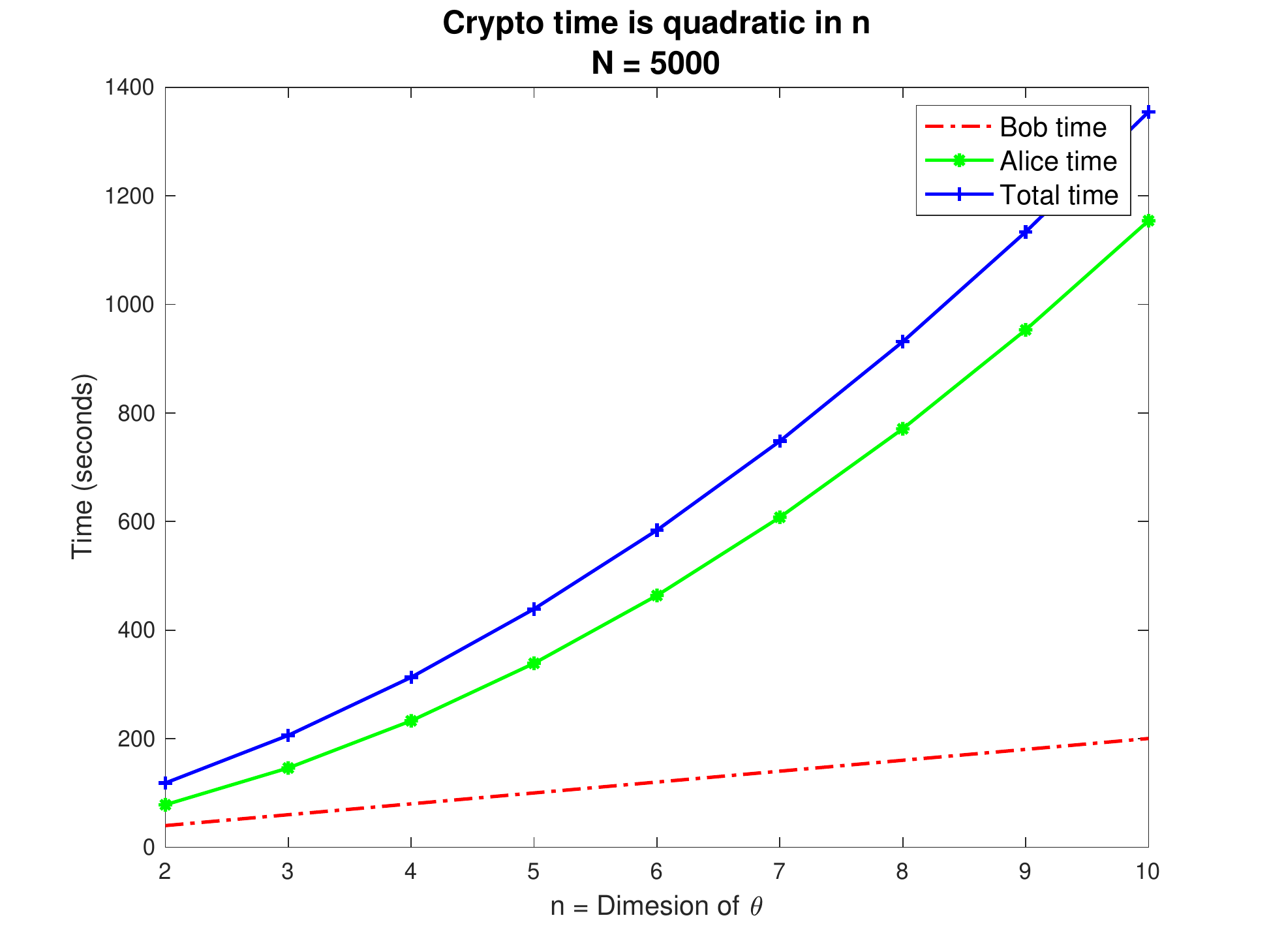}
\caption{Simulation of the total required time of cryptographic routines}
\label{cryptotime}
\end{figure}

Each round of the privacy preserving protocol (starting and ending at \textbf{(1)} in Fig. \ref{onestep}) mainly depends on $n$ and requires: 
\begin{itemize} 
\item Alice time: 
\begin{enumerate}
    \item time to multiply encrypted matrix of size $ n\times n$ by a plain text vector of size $n$ $\implies n^2 (\text{ME} + \text{MM}) $
    \item time to flip the encrypted $X^TY$ and add it to the encrypted result: $ \implies n (\text{MM} + \text{MI}) $
    \item time to encrypt a random vector of size $n$ and add it to the result $ \implies n (\text{E}+\text{MM}) $
    \item sub-protocol time: decrypt an $n$ vector $\implies  n \text{D} $
\end{enumerate}
$$ 
\text{time}^\text{round}_\text{Alice} \approx  n^2 (\text{ME} + \text{MM}) + n (2\text{MM} + \text{MI}+ \text{E} + \text{D})$$
\item Bob time 
\begin{enumerate}
    \item decrypt an $n$ vector,
    \item sub-protocol time: encrypt an $n$ vector and add to plaintext vector,
\end{enumerate} 
$$ \text{time}^\text{round}_\text{Bob} \approx n (\text{D} + \text{E} + \text{MM}) $$
\end{itemize}
Note that we have neglected non-cryptographic operations time with respect to cryptographic routines time. 
In addition, we account for the setup time apart from key generation and parameter negotiation. This time is required only once: 
\begin{align*}
\text{time}^\text{setup}_{Bob} & \equiv n^2 \text{E}  + n(\text{E} +\text{D}) \\  \text{time}^\text{setup}_{Alice} & \equiv n^2 (\text{ME} + 2 \text{MM}) + n (\text{MI}+  3\text{MM} )  
\end{align*}
As shown in Fig. \ref{cryptotime}, the total time cost of the protocol is: 
$$ \text{time}_\text{total} = \text{time}^\text{setup} + N \times \text{time}^\text{round}
$$

\subsection{Communication cost} 
 The protocol requires one round for setup and two rounds of communications per iteration. The exchanged size is $O(n^2)$ and is independent of $m$. This is an advantage since in linear least squares $n$ is usually small and $m$ is large.
\section{Conclusion} 
In this paper, we extended recent research in private machine learning and proposed a practical scheme for secure two party computation of linear regression and linear least squares. Our protocol is based on a fixed point encoding scheme, and masking through one-time random pads for hiding intermediate results. It requires two rounds of communication per step of gradient descent. We presented numerical experiments and simulation results. In future work we aim at implementing our approach with the help of existing secure multi-party computation languages and libraries. The implementation will allow comparing to other approaches from the literature in terms of security, cost and performance. In case where our approach is revealed superior, we estimate that similar multi-party private protocols can be developed for linear means classifier, Fisher's Linear Discriminant Classifier and linear perceptrons.

\section*{Acknowledgments}
The author would like to thank Haitham Bou Ammar for his ideas and the early discussion of the problematic and possible contributions. Thanks also go to Qais Humeid who helped with the Matlab code.
\bibliographystyle{IEEEtran/IEEEtran} 
\bibliography{main}

\end{document}